\documentclass[twocolumn,prb]{revtex4}
\usepackage{graphicx}
\usepackage{dcolumn}
\usepackage{bm}
\usepackage{amsmath}

\begin{document}
\preprint{}
\title{Theory of charge transport in ferromagnetic semiconductor/$s$-wave superconductor junction}
\author{Yoshihiro Mizuno$^{1}$, Takehito Yokoyama$^{2}$, and Yukio Tanaka$^{1}$ }
\affiliation{$^{1}$Department of Applied Physics, Nagoya University, Nagoya 464-8603, Japan. \\
$^2$Department of Applied Physics, University of Tokyo, Tokyo 113-8656, Japan}
\date{\today}

\begin{abstract}
  We study tunneling conductance in ferromagnetic semiconductor/insulator/$s$-wave superconductor junction where Rashba
spin-orbit interaction (RSOI) and exchange field are taken into account in the ferromagnetic semiconductor. We show that normalized conductance at zero voltage has a maximum as a function of RSOI for high transparent interface and finite exchange field. This is because Andreev reflection probability shows a nonmonotonic dependence on RSOI in the presence of the exchange field. On the other hand, for intermediate transparent interface, normalized conductance at zero voltage has a reentrant shape at zero or small exchange field with increasing RSOI but is monotonically increasing by RSOI at large exchange field.
\end{abstract}

\maketitle



%

%



\section{Introduction}
 Spintronics aims to utilize not only charge but also  spin degree of freedom of electrons in electronic devices and circuits.\cite{Zutic,tserkovnyak_rmp_05,brataas_physrep_06,Maekawa,Tatara,Nagaosarev} Electrical spin injection in normal metal is usually achieved by driving a current through ferromagnet/normal metal junction. Recently, spin orbit interaction (SOI) in metal and semiconductors has attracted significant attention in the field of spintronics, since it allows for electrical control of spin without the use of ferromagnets or a magnetic field. Much attention has been paid to the study of the effect of Rashba spin-orbit interaction (RSOI)\cite{Rashba} on transport properties of two dimensional electron gas (2DEG)\cite{DP,Nitta,Hirsch,Governale,Streda,Mishchenko,Schliemann,Sinova,Srisongmuang,Tsung-Wei} because it offers the opportunity of controlling the RSOI and hence spin transport by an external electric field. \cite{Hirsch,Governale,Streda,Mishchenko,Schliemann,Murakami,Sinova,Edelstein,Inoue,Watson,Kato,Wunderlich} The pioneering work by Datta and Das suggested the way to control the precession of the spins of electrons by the RSOI \cite{Rashba} in F/2DEG/F junction (F: ferromagnet) \cite{Datta}. This spin-orbit coupling depends on the applied electric field and can be tuned by a gate voltage.

 There is an attempt to study spintronics in superconducting junction.\cite{Yokoyama,Bai,Feng,Zhang} Charge transports in two dimensional electron gas / $s$-wave superconductor junction with a RSOI has been studied in Ref.\cite{Yokoyama}. It is clarified that for low insulating barrier the tunneling conductance is suppressed by the RSOI while for high insulating barrier it is almost independent of the RSOI. It is also found that the reentrant behavior of the conductance appears at zero voltage as a function of RSOI for intermediate insulating barrier strength. On the other hand, spin dependent transport in ferromagnet / $s$-wave superconductor (F/S) junction is also an important subject in the field of spintronics\cite{Bergeret,Buzdin}. Charge transport in F/S junction also has been studied so far\cite{FS}. The Andreev reflection (AR) in this junction is suppressed because the retro-reflectivity is broken by  the exchange field in the F layer\cite{de Jong}. As a result, the conductance of the junction is suppressed.  However, the interplay between RSOI and exchange field \cite{kato}in superconducting junction still remains unexplored.
In this paper, we study charge transport in ferromagnetic semiconductor/spin-singlet $s$-wave superconductor junction, taking into account RSOI and the exchange field simultaneously\cite{Ohno,Jungwirth} and calculate conductance by changing RSOI, exchange field and the height of insulator at the interface. We show that normalized conductance at zero voltage has a maximum as a function of RSOI for high transparent interface and finite exchange field. This is because AR probability shows a nonmonotonic dependence on RSOI in the presence of the exchange field. On the other hand, for intermediate transparent interface, normalized conductance at zero voltage has a reentrant shape at zero or small exchange field with increasing RSOI but is monotonically increasing by RSOI at large exchange field.

\section{Formulation}
\begin{figure}[htb]
\begin{center}
\scalebox{0.4}{
\includegraphics[width=20.0cm,clip]{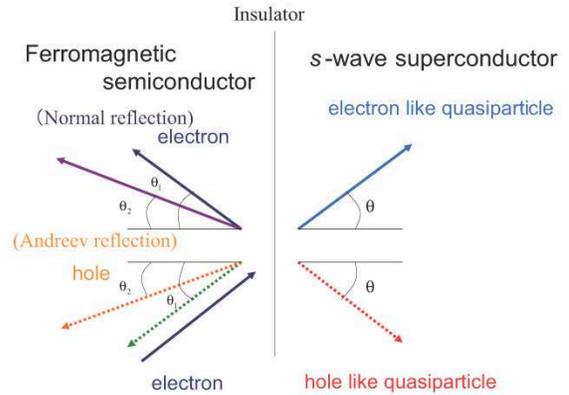}}
\end{center}
\caption{(Color online) Schematic illustration of scattering processes.}
\label{form}
\end{figure}
We consider ballistic ferromagnetic semiconductor/insulator/spin-singlet $s$-wave superconductor (FS/S) junction where the FS/S interface is located at $\it{x}$=0(along y axis), and has infinitely narrow insulating barrier described by the $\delta$ function $\it{U}$$\delta({\it{x}})$, where ferromagnetic semiconductor is modeled as a 2DEG with vertical magnetization.\\
 The effective Hamiltonian including both RSOI and exchange field in spin and Nambu space is given by
\begin{widetext} 
\begin{equation}
\mathcal{H} = \left( {\begin{array}{*{20}c}
   {\xi _k  + U\delta \left( x \right) - H\theta \left( { - x} \right)} & {i\lambda k_ -  \theta \left( { - x} \right)} & 0 & {\Delta \theta \left( x \right)}  \\
   { - i\lambda k_ +  \theta \left( { - x} \right)} & {\xi _k  + U\delta \left( x \right) + H\theta \left( { - x} \right)} & { - \Delta \theta \left( x \right)} & 0  \\
   0 & { - \Delta \theta \left( x \right)} & { - \xi _k  + U\delta \left( x \right) - H\theta \left( { - x} \right)} & { - i\lambda k_ +  \theta \left( { - x} \right)}  \\
   {\Delta \theta \left( x \right)} & 0 & {i\lambda k_ -  \theta \left( { - x} \right)} & { - \xi _k  + U\delta \left( x \right) + H\theta \left( { - x} \right)}  \\
\end{array}} \right)
\end{equation}
\end{widetext}
with $k_ \pm   = k_x  \pm ik_y $, the energy gap $\Delta$, $\xi _k  = \frac{{\hbar ^2 }}{{2m}}\left( {k^2  - k_F^2 } \right)
$,the Fermi wave number ${\it{k}}_{F}$, the exchange field $H$, the Rashba coupling constant $\lambda$, and the step function $\theta(\it{x})$.\\
As shown in Fig. \ref{form}, the wave function $\psi(\it{x})$ for $\it{x}\leq $0 (FS side) is represented using eigen functions of the Hamiltonian:
\begin{widetext}
{\small{
\begin{equation}
\begin{array}{l}\psi(\it{x})=\\e
 ^{ik_y y} \left[ {\frac{1}{{\sqrt 2 }}e^{ik_{1(2)} \cos \theta _{1(2)} x} \alpha _{1(2)} \left( {\begin{array}{*{20}c}
   {i\frac{{\left(  -  \right)\lambda k_{1(2) - } }}{{H + ( - )\sqrt {(\lambda k_{1(2)} )^2  + H^2 } }}}  \\
   1  \\
   0  \\
   0  \\
\end{array}} \right) + \frac{{a_{1(2)} }}{{\sqrt 2 }}e^{ik_1 \cos \theta _1 x} \alpha _1 \left( {\begin{array}{*{20}c}
   0  \\
   0  \\
   {i\frac{{\lambda k_{1 + } }}{{H + \sqrt {(\lambda k_1 )^2  + H^2 } }}}  \\
   1  \\
\end{array}} \right)} \right. \\ 
 \left. { + \frac{{b_{1(2)} }}{{\sqrt 2 }}e^{ik_2 \cos \theta _2 x} \alpha _2 \left( {\begin{array}{*{20}c}
   0  \\
   0  \\
   {i\frac{{\lambda k_{2 + } }}{{H - \sqrt {(\lambda k_2 )^2  + H^2 } }}}  \\
   1  \\
\end{array}} \right) + \frac{{c_{1(2)} }}{{\sqrt 2 }}e^{ - ik_1 \cos \theta _1 x} \alpha _1 \left( {\begin{array}{*{20}c}
   { - i\frac{{\lambda k_{1 + } }}{{H + \sqrt {(\lambda k_1 )^2  + H^2 } }}}  \\
   1  \\
   0  \\
   0  \\
\end{array}} \right) + \frac{{d_{1(2)} }}{{\sqrt 2 }}e^{ - ik_2 \cos \theta _2 x} \alpha _2 \left( {\begin{array}{*{20}c}
   { - i\frac{{\lambda k_{2 + } }}{{H - \sqrt {(\lambda k_2 )^2  + H^2 } }}}  \\
   1  \\
   0  \\
   0  \\
\end{array}} \right)} \right] \\ 
 \end{array}
\end{equation}}}
\end{widetext}
for an injection wave with wave number $\it{k}_{\rm{1(2)}}$ where {{\tiny{$k_{1\left( 2 \right)}  = \sqrt {2\left( {\frac{{m\lambda }}{{\hbar ^2 }}} \right)^2  + k_F^2  +(-) \sqrt {\left\{ {2\left( {\frac{{m\lambda }}{{\hbar ^2 }}} \right)^2  + k_F^2 } \right\}^2  + \left( {\frac{{2mH}}{{\hbar ^2 }}} \right)^2  - k_F^4 } } $}}}, and $k_{1(2) \pm }  = k_{1(2)} \rm{e}^{ \pm i\theta _{1(2)} }$. ${\it{a}}_{1(2)}$ and ${\it{b}}_{1(2)}$ are AR coefficients. ${\it{c}}_{1(2)}$ and ${\it{d}}_{1(2)}$ are normal reflection coefficients.
 $\theta_{1(2)}$ is the angle of the wave number $\it{k}_{\rm{1(2)}}$ with respect to the interface normal, and $
\alpha _{1(2)}$ = $\sqrt {1 + ( - )\frac{H}{{\sqrt {(\lambda k_{1(2)} )^2  + H^2 } }}} 
$.
Note that since the translational symmetry holds for the $\it{y}$ direction, the momenta parallel to the interface are conserved: ${\it{k}}_{\it{y}}={\it{k}}_{{\it{F}}}\sin\theta={\it{k}}_1\sin\theta_1={\it{k}}_2\sin\theta_2$ where $\theta$ denotes the direction of motions of quasiparticles in the S measured from the interface normal.\\
The wave function follows the boundary conditions\cite{Yokoyama,Molenkamp},
\begin{equation}
\begin{array}{l}
 \left. {\psi \left( x \right)} \right|_{x =  + 0}  = \left. {\psi \left( x \right)} \right|_{x =  - 0},  \\ 
 \left. {v_x \psi \left( x \right)} \right|_{x =  + 0}  - \left. {v_x \psi \left( x \right)} \right|_{x =  - 0}  = \frac{\hbar }{{mi}}\frac{{2mU}}{{\hbar ^2 }}\tau_3\psi\left( 0 \right), \\ 
\\
 \tau _3  = \left( {\begin{array}{cccc}
   1 & 0 & 0 & 0  \\
   0 & 1 & 0 & 0  \\
   0 & 0 & { - 1} & 0  \\
   0 & 0 & 0 & { - 1} 
\end{array}} \right).
\end{array}
\end{equation}
According to Ref. \cite{Yokoyama}, we derive a formula for the tunneling conductance, and obtain the dimensionless conductance represented in the form\\
\begin{widetext}
\begin{eqnarray}
 \sigma _S  &=& N_1 \int_{ - \theta _C }^{\theta _C } {} \left[ {{\it{X}} + \left| {a_1 } \right|^2  {\it{X}}+ \left| {b_1 } \right|^2 {\it{Y}}\lambda _{21} }
{ - \left| {c_1 } \right|^2  {\it{X}}- \left| {d_1 } \right|^2 {\it{Y}}\lambda _{21} } \right]\cos \theta d\theta  \nonumber \\ 
  &+& N_2 \int_{ - \frac{\pi }{2}}^{\frac{\pi }{2}} {{\mathop{\rm Re}\nolimits} } \left[ {{\it{Y}} + \left| {a_2 } \right|^2 {\it{X}}\lambda _{12}  + \left| {b_2 } \right|^2 \it{Y}}{ - \left| {c_2 } \right|^2 {\it{X}}\lambda _{12}  - \left| {d_2 } \right|^2{\it{Y}}} \right]\cos \theta d\theta  \nonumber \\ 
  &=& \int_{ - \theta _C }^{\theta _C } {\left[ {1 + \left| {a_1 } \right|^2  + \left| {b_1 } \right|^2 \frac{{\it{Y}}}{{\it{X}}}\lambda _{21} } { - \left| {c_1 } \right|^2  - \left| {d_1 } \right|^2 \frac{{\it{Y}}}{{\it{X}}}\lambda _{21} } \right]\cos \theta d\theta }  \nonumber \\ 
  &+& \int_{ - \frac{\pi }{2}}^{\frac{\pi }{2}} {{\mathop{\rm Re}\nolimits} } \left[ {1 + \left| {a_2 } \right|^2 \frac{{\it{X}}}{{\it{Y}}}\lambda _{12}  + \left| {b_2 } \right|^2 }{ - \left| {c_2 } \right|^2 \frac{{\it{X}}}{{\it{Y}}}\lambda _{12}  - \left| {d_2 } \right|^2 } \right]\cos \theta d\theta  \\ 
  &\equiv& \left[ {1 + A_1  + B_1 } \right.\left. { - C_1  - D_1 } \right]\int_{ - \theta _C }^{\theta _C } {\cos \theta d\theta }  + 2\left( {1 + A_2  + B_2  - C_2  - D_2 } \right) \\ &\equiv& \sigma_{\rm{S1}}+\sigma_{\rm{S2}},
\end{eqnarray}
\end{widetext}
where we define ${\it{X}}=\left( {1 + \frac{{m\lambda ^2 }}{{\hbar ^2 \sqrt {(\lambda k_1 )^2  + H^2 } }}} \right)$ and ${\it{Y}}=\left( {1 - \frac{{m\lambda ^2 }}{{\hbar ^2 \sqrt {(\lambda k_2 )^2  + H^2 } }}} \right)$.
$A_1$, $A_2$, $B_1$ and $B_2$ denote Andreev reflection probability, 
while  $C_1$, $C_2$, $D_1$ and $D_2$ denote normal reflection probability.
In the above, 
${\it{N}}_{1}$ and ${\it{N}}_{2}$ are defined as the densities of states normalized by those with $\lambda$=0 and $\it{H}$=0 for wave numbers ${\it{k}}_1$ and  ${\it{k}}_2$, respectively:
\begin{equation}
N_{1\left( 2 \right)}  = \frac{1}{{1 + \left(  -  \right)\frac{{m\lambda ^2 }}{{\hbar ^2 \sqrt {H^2  + (\lambda k_{1\left( 2 \right)})^2 } }}}}.
\end{equation}
$\lambda_{12}$ and $\lambda_{21}$ are defined as the following:
\begin{equation}
\lambda _{12}  = \frac{{k_1 \cos \theta _1 }}{{k_2 \cos \theta _2 }}_, \lambda _{21}  = \frac{{k_2 \cos \theta _2 }}{{k_1 \cos \theta _1 }}.
\end{equation}
The critical angle $\theta_{c}$ is defined as $\cos\theta_{c}=\sqrt{1-(\frac{{\it{k}}_{1}}{{\it{k}}_{F}})^{2}}$.
${\sigma}_{\rm{N}}$ is given by the conductance for normal states, i.e., ${\sigma}_{\rm{s}}$ for $\Delta=0$. We define the normalized conductance as ${\sigma}_{\rm{T}}$=$\frac{{\sigma}_{\rm{S}}}{{\sigma}_{\rm{N}}(\beta=0)}$ (or $\frac{{\sigma}_{\rm{S}}}{{\sigma}_{\rm{N}}(\gamma=0)})$ as a function of $\beta$ (or $\gamma$) and the parameters as $\beta=\frac{2\it{m}\lambda}{\hbar^{2}\it{k}_{F}}$, $\gamma=\frac{2\it{mH}}{\hbar^2{\it{k}}^{2}_{F}}$ and ${\it{Z}}=\frac{2{\it{m}}{\it{U}}}{\hbar ^2 k _F}$. Similarly, we define  the normalized conductance as ${\sigma}_{\rm{1(2)}}$=$\frac{{\sigma}_{\rm{S1(2)}}}{{{\sigma}_{\rm{N}}}(\beta=0)}$ (or $\frac{{\sigma}_{\rm{S1(2)}}}{{{\sigma}_{\rm{N}}}(\gamma=0)})$. We neglect the difference of effective masses between FS and S in the present paper since it is effectively renormalized into the barrier parameter $Z$. 


\section{Results}
First, we study the normalized tunneling conductance $\sigma_{\rm{T}}$ as a function of RSOI at zero voltage (${\it{eV}}=0$). For Z=0 (Fig. \ref{z0v0be}(a)) where the AR probability is high, normalized conductance ${\sigma}_{\rm{T}}$ at zero voltage has a maximal value as a function of RSOI for finite $\gamma$.  This can be understood by decomposing the conductance into two parts, ${\sigma}_{1}$ and ${\sigma}_{2}$(Fig. \ref{z0v0be}(b)):
\begin{figure}[htb]
\begin{center}
\scalebox{0.4}{
\includegraphics[width=20.0cm,clip]{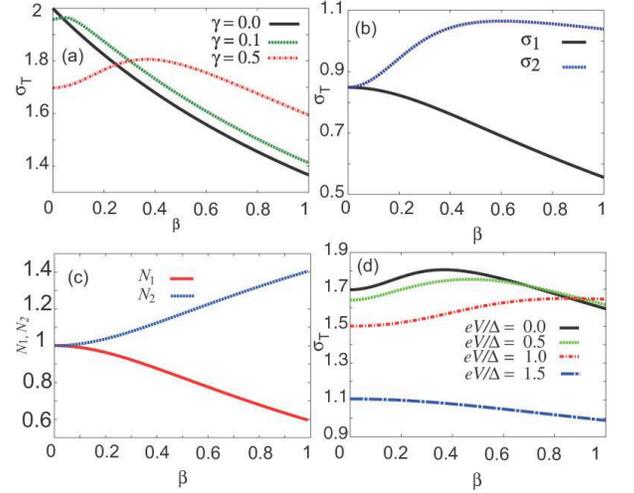}}
\end{center}
\caption{(color online) (a) Normalized tunneling conductance 
($\sigma_{\rm{T}}=\frac{{\sigma}_{\rm{S}}}{{\sigma}_{\rm{N}}(\beta=0)}$) 
as a function of $\beta$  at ${\it{Z}}$=0 and zero voltage with  $\gamma$=0,0.1,0.5. (b)${\sigma}_{1}$(injection with ${\it{k}}_{1})$ and ${\sigma}_{2}$(injection with ${\it{k}}_{2})$ at zero voltage for $\gamma$=0.5.(c)${\it{N}}_{1}$ and ${\it{N}}_{2}$ are the densities of states normalized by those with $\lambda$=0 and $\it{H}$=0 for wave numbers ${\it{k}}_1$ and  ${\it{k}}_2$, respectively.  (d)  Normalized tunneling conductance at non-zero voltage and zero voltage for $\gamma$=0.5 with  $\it{eV}/\Delta$=0,0.5,1.0, and 1.5. }
\label{z0v0be}
\end{figure}
\begin{figure}[htb]
\begin{center}
\scalebox{0.4}{
\includegraphics[width=20.0cm,clip]{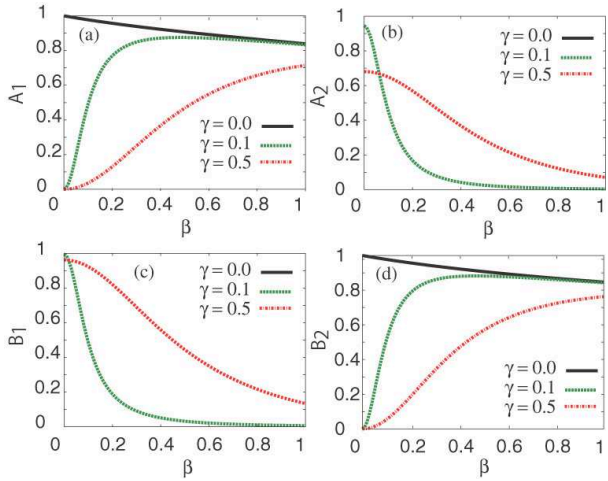}}
\end{center}
\caption{(color online) The angular averaged AR probability as a function of $\beta$ at zero voltage for $\it{Z}$=0  with  $\gamma$=0,0.1,0.5.  At $\gamma=0$, $A_2=B_1=0$.}
\label{ARz0v0be}
\end{figure}
${\sigma}_{1}(=\sigma_{\rm{S}}/\sigma_{\rm{N}}(\beta=0))$, which stems from the injection of wave function with ${\it{k}}_{1}$, is reduced by an increase of RSOI but ${\sigma}_{2}(=\sigma_{\rm{S}}/\sigma_{\rm{N}}(\beta=0))$, which originates from the injection of wave function with ${\it{k}}_{2}$, increases by an increase of RSOI as shown in Fig. \ref{z0v0be}(b). 
These features can be explained by the AR probabilities as shown in Fig. \ref{ARz0v0be}. We find that the difference between ${\sigma}_{1}$ and ${\sigma}_{2}$ stems from increasing and decreasing AR probabilities: ${\it{A}}_{1}$ is increasing (Fig. \ref{ARz0v0be}(a)) while ${\it{B}}_{1}$ is reduced by RSOI (Fig. \ref{ARz0v0be}(c)). Then, the suppression dominates the enhancement. Therefore,  ${\sigma}_{1}$ is reduced by increasing of RSOI. On the other hand, ${\it{A}}_{2}$ is reduced (Fig. \ref{ARz0v0be}(b)) and
${\it{B}}_{2}$ is increasing by RSOI (Fig. \ref{ARz0v0be}(d)). The enhancement dominates the suppression for $0< \beta <0.6$ but the suppression dominates the enhancement for $0.6 < \beta < 1.0$. Therefore, ${\sigma}_{2}$ is enhanced by increasing of RSOI. For both cases, the same band AR (${\it{A}}_{1}$,${\it{B}}_{2}$) is increasing. 
For $\beta$=0, ${\it{k}}_{1}$ (${\it{k}}_{2}$) corresponds to down (up) spin band in the ferromagnet. In this case,  AR within the same band is forbidden because electrons with the same spin do not form Cooper pairs in singlet $s$-wave superconductor. However, for $\beta\neq0$, RSOI causes spin mixing and hence the state characterized by ${\it{k}}_{1}$(${\it{k}}_{2}$) consists of up and down spin states as increasing RSOI. For this reason, the same band AR becomes possible. On the other hand, the interband AR (${\it{A}}_{2}$, ${\it{B}}_{1}$) is reduced by RSOI because the normalized density of states ${\it{N}}_{1(2)}$ is reduced (increasing) as shown in  Fig. \ref{z0v0be}(c). In this way, we understand that the competition between these two contributions 
causes the non-monotonous dependence of $\sigma_{\rm{T}}$ on RSOI. 
This feature is also seen at non-zero voltage below the gap as shown in Fig. \ref{z0v0be}(d). 

\begin{figure}[htb]
\begin{center}
\scalebox{0.4}{
\includegraphics[width=20.0cm,clip]{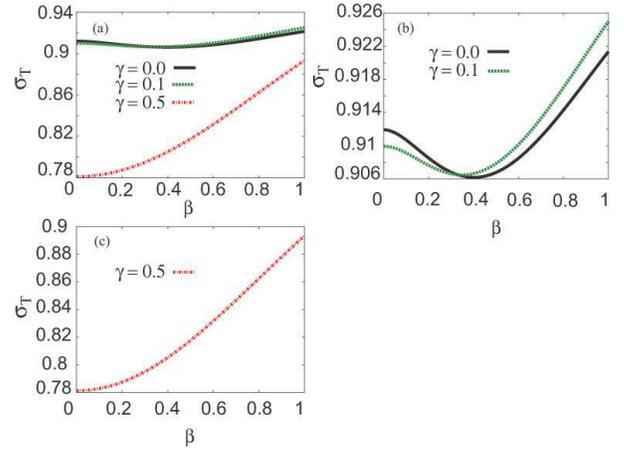}}
\end{center}
\caption{(color online) Normalized tunneling conductance as a function of $\beta$ ($\frac{{\sigma}_{\rm{S}}}{{\sigma}_{\rm{N}}(\beta=0)}$) at ${\it{Z}}$=1.0  with  $\gamma$=0,0.1,0.5 at zero voltage.(a)Normalized tunneling conductance with  $\gamma$=0,0.1,0.5. (b)$\gamma$=0 and 0.1(c)$\gamma$=0.5.}
\label{z1be}
\end{figure}

\begin{figure}[htb]
\begin{center}
\scalebox{0.4}{
\includegraphics[width=20.0cm,clip]{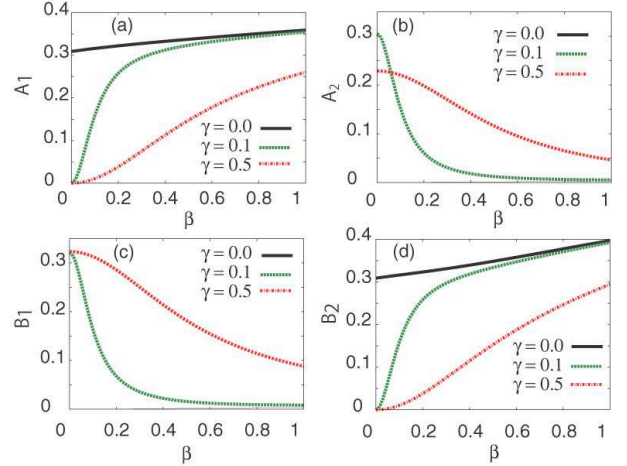}}
\end{center}
\caption{(color online) The angular averaged AR probability at zero voltage for $\it{Z}$=1.0  with $\gamma$=0,0.1,0.5. At $\gamma=0$, $A_2=B_1=0$.}
\label{fig4}
\end{figure}

 For $\it{Z}=$1.0 (Fig. \ref{z1be}), $\sigma_{\rm{T}}$ has a reentrant shape at zero or small $\gamma$(i.e., $\gamma$=0.1) with increasing RSOI(Fig. \ref{z1be}(b)). This result is consistent with the previous work\cite{Yokoyama}. Meanwhile, $\sigma_{\rm{T}}$ is monotonically increasing with RSOI at large $\gamma$(i.e., $\gamma$=0.5) (Fig. \ref{z1be}(c)) at zero voltage. These features can be understood in a way similar to those at $Z=0$.
From Fig. \ref{fig4}, we also find that the AR probabilities for $\it{Z}=$1.0 (Fig. \ref{fig4}) are similar to those for $\it{Z}=$0 except that their magnitudes are small because of  the intermediate barrier strength $\it{Z}=$1.0. Hence, the behavior of $\sigma_{\rm{T}}$ in Fig. \ref{z1be}(c) is due to the fact that the enhanced AR probabilities overcome those reduced, resulting in the enhancement of the conductance.

%

\begin{figure}[htb]
\begin{center}
\scalebox{0.4}{
\includegraphics[width=20.0cm,clip]{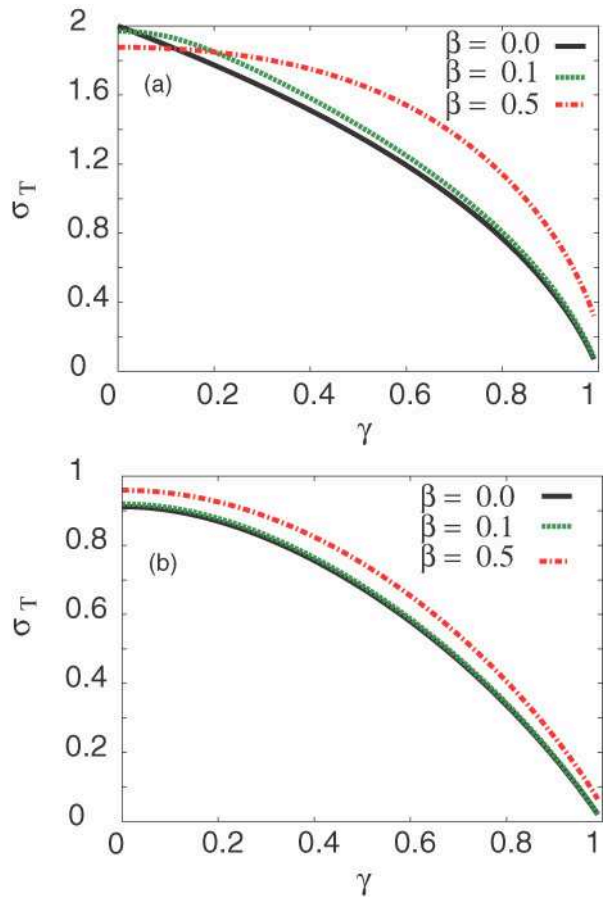}}
\end{center}
\caption{(color online) Normalized tunneling conductance as a function of $\gamma$ ($\frac{{\sigma}_{\rm{S}}}{{\sigma}_{\rm{N}}(\gamma=0)}$) at zero voltage with  $\beta$=0,0.1,0.5. (a) ${\it{Z}}$=0. (b) ${\it{Z}}$=1.0. }
\label{con-gaz0}
\end{figure}

Next, we show $\sigma_{\rm{T}}$ as a function of $\gamma$ at zero voltage.
For $Z=0$ (Fig.\ref{con-gaz0}(a)) where AR probability is high, conductance is reduced by exchange field. For $\it{Z}$=1.0 where AR probability is intermediate, ${\sigma}_{\rm{T}}$ is monotonically decreasing with increasing $\gamma$ as seen in Fig.\ref{con-gaz0}(b). This is because Cooper pairs are broken by the exchange field.\cite{de Jong}

\section{Conclusions}

In the present paper, we have studied the tunneling conductance in ferromagnetic semiconductor/insulator/$s$-wave superconductor junction with RSOI and exchange field. For high transparent interface, we showed that normalized conductance $\sigma_{\rm{T}}$ at zero voltage has a maximal value as a function of RSOI for finite exchange field. Because RSOI makes spin polarized states by exchange field mixture of spin up and down states, Andreev reflection probability shows a nonmonotonic dependence on RSOI in the presence of the exchange field, which leads to the nonmonotonic behavior of the conductance. We also clarified that normalized conductance has a reentrant shape at zero or small exchange field with increasing RSOI but normalized conductance $\sigma_{\rm{T}}$ is monotonically increasing by RSOI at large exchange field for intermediate transparent interface.  It is also found that $\sigma_{\rm{T}}$ as a function of exchange field at zero voltage is reduced by exchange field. We hope that the results obtained are useful for a better understanding of related topic and experiments.

There are several future works. 
In the present paper, we focus on the conventional $s$-wave superconductor. 
For unconventional superconductor, it is known that 
the Andreev bound state is formed at the interface \cite{ABS}. 
It is also an interesting issue to study ferromagnetic semiconductor / 
unconventional superconductor junctions. 
Beside this problem, to investigate symmetry of Cooper pair is a challenging issue. 
It has been established that odd-frequency pairing amplitude is induced in the 
normal metal / superconductor junction due to the breakdown of the 
translational symmetry\cite{oddfrequency}. 
It is also challenging to clarify the pairing amplitude in the ferromagnetic semiconductor region in ferromagnetic semiconductor / superconductor junctions. 
Also, spin transport in superconducting junctions is an important problem. 
Spin conductance in the non-centrosymmetric superconductor 
has been studied in Ref.\cite{Nagaosa}. It is very interesting to extend the present approach by including non-centrosymmetric superconductors. \cite{Yokoyama2,Iniotakis,Linder,Wu}

\section{Acknowledgments}
The authors appreciate useful and fruitful discussions with S. Onari and J. Inoue. This study has been supported by Grants-in-Aid for the 21st Century COEeFrontiers of Computational Science' and the JSPS (T.Y.).
%

\end{document}